\documentclass[aps, a4paper,superscriptaddress, nofootinbib, twocolumn]{revtex4}
\usepackage{epsfig}
\usepackage{graphicx}
\usepackage{amsmath}
\usepackage{color}
\newcommand{\be}{\begin{equation}}
\newcommand{\ee}{\end{equation}}
\newcommand{\beq}{\begin{equation}}
\newcommand{\eeq}{\end{equation}}
\newcommand{\bea}{\begin{eqnarray}}
\newcommand{\eea}{\end{eqnarray}}

\newcommand{\ie}{{i.e.}}

\newcommand{\gev}{\, \text{GeV}}

\newcommand{\no}{\notag \\}
\newcommand{\fs}{\,.}
\newcommand{\co}{\,,}
\newcommand{\D}{\displaystyle}

\newcommand{\indL}{L}
\newcommand{\indT}{T}
\newcommand{\indP}{P}
\newcommand{\inel}{\text{inel}}
\newcommand{\el}{\text{el}}

\newcommand{\nuth}{\nu_\text{th}}

\newcommand{\subsec}{subsection}

\newcommand{\alphaem}{\alpha_\text{em}}
\newcommand{\mN}{m}
\newcommand{\mpi}{M_\pi}

\renewcommand{\Im}{\text{Im}\,}
\newcommand{\thr}{\text{th}}
\newcommand{\Regge}{\text{Regge}}
\newcommand{\bsp}{\begin{sloppypar}}
\newcommand{\esp}{\end{sloppypar}}

\begin{document}



\title{\Large Constraints on the virtual Compton scattering on the nucleon in a new dispersive formalism}
\author{Irinel Caprini}
\affiliation{Horia Hulubei National Institute for Physics and Nuclear Engineering, POB MG-6, 077125 Bucharest-Magurele, Romania}

\begin{abstract}\vspace{0.5cm}{The dispersive representation of
the virtual Compton forward scattering amplitude  has been recently reexamined in connection with the evaluation of the Cottingham formula for the proton-neutron electromagnetic mass difference and the proton radius puzzle.  The most difficult part of the analysis is related to one of the invariant amplitudes, denoted as  $T_1(\nu, Q^2)$, which requires a subtraction in the standard dispersion relation  with respect to the energy $\nu$ at fixed photon momentum squared $q^2=-Q^2$.  We propose an alternative dispersive framework, which implements analyticity and unitarity by combining the Cauchy integral relation at low and moderate energies with the modulus representation of the amplitude at high energies. Using techniques of functional analysis, we derive a necessary and sufficient condition for the consistency with analyticity of the subtraction function $S_1(Q^2)=T_1(0, Q^2)$, the cross sections measured at low and moderate energies and the Regge model assumed to be valid at high energies. From this condition we obtain model-independent constraints on the  subtraction function, confronting them with the available information on nucleon magnetic polarizabilities and results reported recently in the literature.  The formalism can be used also for testing the existence of a fixed pole at $J=0$ in the angular momentum plane, but more accurate data are necessary for a definite answer.}
\end{abstract}
\maketitle

\section{Introduction}

The virtual Compton forward scattering on the nucleon is of interest for the calculation of the proton-neutron electromagnetic mass difference by the Cottingham formula \cite{Cottingham, Collins}. 
An early dispersive evaluation of this formula, performed in \cite{GL1975}, was updated recently in  \cite{GHLR2015} and the problem was examined also in  other recent  papers \cite{WalkerLoud:2012bg}-\cite{Erben:2014hza}. In addition, the virtual Compton  scattering is of interest for the proton radius puzzle (for a review see \cite{Carlson:2015}). It has been considered also by several authors, in particular in the context of Finite Energy Sum Rules (FESR), in connection with the topic of a fixed $J=0$ pole in the amplitude \cite{Creutz, DaGi, Brodsky:2008qu, Gorc1, Muller:2015vha}.

The nucleon matrix element of the time-ordered product, $\langle p|Tj^\mu(x)j^\nu(y)|p\rangle$, where $j^\mu$ is the electromagnetic current, is described in terms of two  invariant amplitudes $T_i(\nu,Q^2)$, $i=1,2$, free of kinematical singularities and zeros, which depend on  $\nu=p\cdot q/m$ and $Q^2=-q^2$, where $p$ and $q$ denote the nucleon and photon momenta and $m$ the nucleon mass.  The functions $T_i(\nu,Q^2)$ at fixed $Q^2\ge 0$ are real analytic in the $\nu$ plane cut along $\nu\ge\nu_0$ and $\nu\le-\nu_0$, where $\nu_0$ is the lowest unitarity threshold. Following \cite{GHLR2015}, we denote the structure functions as  $V_i(\nu,Q^2)$:
\beq
\Im T_i(\nu+i \epsilon,Q^2)=\pi V_i(\nu,Q^2)\co~~\nu\geq \nu_0\co\,~ Q^2\ge 0.
\eeq
Due to the contributions arising from Regge exchanges which exhibit a growth at infinity,  $V_1(\nu,Q^2) \sim \nu^\alpha$ with $\alpha>0$, the standard dispersion relation for $T_1(\nu, Q^2)$ as a function of $\nu$ at fixed $Q^2$ requires a subtraction, while for $T_2(\nu, Q^2)$ an unsubtracted relation is valid. 

 Choosing the subtraction point at $\nu=0$ and denoting
\be 
S_1(Q^2)\equiv T_1(0,Q^2),\label{eq:S}
\ee
 the dispersion relation satisfied by the amplitude $T_1$ at fixed $Q^2$ is written in the form \cite{GHLR2015}
\beq\label{eq:DR}
T_1(\nu,Q^2)= S_1(Q^2)+2\nu^2\int_0^\infty \frac{d\nu'}{\nu'} \,\frac{V_1(\nu',Q^2)}{\nu'^2-\nu^2-i\epsilon},
\eeq
taking into account the fact that the amplitude is an even function of $\nu$ at fixed $Q^2$. 
The dispersive integral  can be evaluated using the structure function $V_1(\nu,Q^2)$ expressed in terms of the electron-proton cross sections measured at low and moderate energies and parametrized by Regge model at high energies. 

The  problem considered in this work is the determination of the subtraction function $S_1(Q^2)$.
We note that at $Q^2=0$ the subtraction function is expressed in terms of the nucleon polarizabilities.
The view adopted in Refs. \cite{WalkerLoud:2012bg}-\cite{Erben:2014hza} is that the subtraction 
function  $S_1(Q^2)$ for  nonzero values of $Q^2$ cannot be calculated in terms of physical observables and represents an arbitrary input.   Accordingly, specific models for  $S_1(Q^2)$ have been adopted in these references in order to evaluate the dispersion relation (\ref{eq:DR}).  Combined with the poor experimental knowledge of the magnetic polarizability of the neutron \cite{Griesshammer},  the arbitrary $Q^2$ dependence of the subtraction function  represents the main source of uncertainty in the estimate of the proton-neutron
mass difference obtained in these references.

Of course, the subtraction function in a dispersion relation is arbitrary if the available  information refers exclusively to the discontinuity of the amplitude across the cut (which for real-analytic functions is related to the imaginary part). However, in the present case more information is available on the amplitude at large energies in the frame of Regge model.
In Refs. \cite{GL1975, GHLR2015} it was shown that, if  the amplitude $T_1$ at fixed $Q^2$ does not have a fixed pole at $J=0$  in the angular momentum plane, the subtraction function itself can be calculated in terms of the physical electroproduction cross sections. The detailed analysis performed in \cite {GHLR2015}  shows that the results obtained from this assumption for the difference of the proton and neutron polarizabilities are consistent with experiment and somewhat more precise, which may be viewed as a nontrivial test of the hypothesis that the Compton amplitude is free of fixed poles. 

In the present work we develop an alternative dispersive framework for the investigation of the subtraction 
function in the virtual Compton scattering, in which no explicit assumption about the presence or the absence of the fixed poles is made.  We start from the remark that the Regge model predicts  at large energies not only the imaginary
part of the amplitude,  but the amplitude itself.  The  high-energy behavior of the amplitude is of the standard  Regge asymptotic form \cite{Collins:1977jy, CCL}
\be
T(s,t) \sim - \frac{\pi \beta_{\alpha}(t)}{\sin\pi \alpha(t) } 
\{\exp[-i\pi\alpha(t)]+\tau\}s^{\alpha(t)}\co
\label{eq:T1RStandard}
\ee
where $s$ is the c.m energy squared (connected to the variable $\nu$ by $s= 2 m\nu+m^2-Q^2$) and $t$ the momentum transfer (in the present case $t=0$ and the signature $\tau=1$). Therefore, from the Regge parametrization of the imaginary part of the amplitude it is possible to recover also the real part, which is in a particular relation to it, as follows from (\ref{eq:T1RStandard}).

If one knows both the real and imaginary parts of a function along a part of the unitarity cut, the function can be predicted in principle  everywhere in the complex plane by the uniqueness of analytic continuation. However, it is known that in practice this task is impossible, due to the fact that analytic continuation is an ill-posed problem in the Hadamard sense \cite{Hadamard}: this means that small uncertainties of the input are amplified in an uncontrolled way outside the original domain. The analytic continuation can be nevertheless ``stabilized" in some cases, for instance if one restricts the class of admissible functions to a compact set.  Such a condition is provided, for instance, by  the knowledge of the modulus of the function along the whole boundary  (see \cite{Ciulli} and references therein).  Guided by these ideas, in the present work we  develop a formalism which uses  as input at high energies the modulus of the amplitude, provided by the  Regge model.

It is useful to recall that in the modulus representation,  where one
constructs an analytic function starting from its modulus on the boundary of the analyticity domain
rather than from its discontinuity (imaginary part), much weaker assumptions at infinity
are needed (since the integral representation contains the logarithm of the input
modulus). As a consequence, if one assumes that the modulus of a function is known along the boundary, it is possible to derive rigorous upper and lower bounds on the values of the function at all points  inside the holomorphy domain, in particular at $\nu=0$. Therefore, although  the subtraction function cannot be calculated exactly, its value is not completely arbitrary. 

 In the present paper we propose a dispersive formalism which uses the Cauchy integral relation to account for the knowledge of the imaginary part on a part of the cut, and the modulus representation on the remaining part. As we shall see, in general this input does not determine uniquely the amplitude, but predicts a whole class of ``admissible" functions to which the physical amplitude belongs. By means of the solution of a functional extremal problem on this class, it is possible to formulate a rigorous necessary and sufficient condition for the  consistency of the input given on the boundary with the values of the amplitude inside the holomorphy domain. This can be exploited to constrain the subtraction function and to investigate the possible fixed poles in the Regge asymptotic behavior. The method can be applied to both invariant amplitudes, however we restrict in this paper to the amplitude $T_1$.  We note that the method is  applicable to a general Regge behavior,
 including the Pomeron contribution, allowing therefore the separate discussion of the proton and neutron amplitudes. 

  The paper is organized as follows: in the next section we review the physical information  used as input, consisting of  the structure function at low and moderate energies and the Regge model at high energies. In Sec. \ref{sec:optimal} we formulate an extremal problem whose solution provides a necessary and sufficient condition for the consistency with analyticity of the cross sections, the Regge model and the subtraction function. The solution of this problem is given in the Appendix, while in Sec. \ref{sec:cn} we discuss the numerical evaluation of the quantities entering the solution. In Sec. \ref{sec:applic} we present the results of our analysis: we first discuss the consistency of the input at $Q^2=0$, where the subtraction function is related to the magnetic polarizability. We derive then parametrization-free upper and lower bounds on the subtraction function  $S_{1}(Q^2)$  and compare them with the results obtained recently in \cite{GHLR2015} and with the parametrization proposed in \cite{Erben:2014hza}.  Section \ref{sec:summary} summarizes our main results and conclusions.
 
\section{Physical input \label{sec:formulation}}

 We consider the so-called ``inelastic" amplitude defined as in  \cite{GHLR2015} by subtracting from the total amplitude the elastic contribution
\beq\label{eq:inel}
T_1^\inel(\nu^2, Q^2)= T_1(\nu^2,Q^2)-T_1^\el(\nu^2,Q^2),
\eeq
where  $T_1^\el$  has a simple expression \cite{GHLR2015} 
\begin{align}
 T_1^\el(\nu^2,Q^2)= -\frac{4\mN^2Q^2\, [G_E^2(Q^2)-G_M^2(Q^2)]}{(4\mN^2\nu^2-Q^4)(4\mN^2+Q^2)}
\label{eq:Tel} 
\end{align}
in terms of the Sachs electromagnetic  form factors of the nucleon, $G_E(t)$ and $G_M(t)$.
 In (\ref{eq:inel}), the notation  emphasizes the fact that the amplitudes, being even functions of $\nu$,  depend actually on $\nu^2$.  At fixed $Q^2$, the amplitude  $T_1^\inel(\nu^2, Q^2)$ is a real-analytic  function in the $\nu^2$-complex plane, cut along the real axis above  $\nuth^2$, where $\nuth= \mpi+(\mpi^2-Q^2)/2\mN$ is the threshold due to the $\pi N$ intermediate state. The discontinuity across the cut is related to the imaginary part
\beq
\Im T_1^\inel(\nu^2+i \epsilon,Q^2)=\pi V_1(\nu,Q^2), ~~\nuth^2\leq\nu^2\leq\nu^2_h, \label{eq:c1}
\eeq
where the structure function $V_1$ is expressed in terms of the longitudinal $\sigma_{\indL}$  and transversal $\sigma_{\indT}$ electroproduction cross sections as  \cite{GHLR2015}
\begin{align}
 V_1(\nu,Q^2)&=  \frac{\mN\nu}{2\alphaem}k(\nu,Q^2)\big\{\bar{\sigma}_{\indL}(\nu,Q^2)-\sigma_{\indT}(\nu,Q^2)\big\}\co\no
\bar{\sigma}_{\indL}(\nu,Q^2)&\equiv \frac{\nu^2}{Q^2}\sigma_{\indL}(\nu,Q^2)\co\no
k(\nu,Q^2)&\equiv\frac{1}{2\pi^2} \frac{\nu -Q^2/2\mN}{\nu (\nu^2+Q^2)}\fs
\label{eq:VsigmaLT} 
\end{align}

In our treatment we shall assume that $V_1(\nu,Q^2)$ is experimentally accessible from the cross sections measured for $Q^2\geq 0$ below a certain high-energy value of $\nu^2$, denoted as $\nu^2_h$.

At higher energies, $T^\inel_1(\nu^2,Q^2)$ is approximately described by Regge model.  In the standard dispersion relations, only the structure function $V_1(\nu,Q^2)$ is obtained from the Regge parametrization. However, the Regge model predicts also the real part, which is related to the imaginary part through the standard form (\ref{eq:T1RStandard}). Therefore, we shall assume that the  modulus of the amplitude along the cut above $\nu^2_h$ satisfies the condition
\beq
 |T^\inel_1(\nu^2,Q^2)|= |T_1^{\Regge}(\nu^2,Q^2)|,~~ \nu^2 \ge \nu^2_h,\label{eq:c2}
\eeq
where $T_1^{\Regge}(\nu^2,Q^2)$ denotes the dominant Regge contributions, given by the Pomeron $P$ and the  degenerate $f$ and $a_2$ trajectories.  From the arguments given in the next section, it will be clear that the bounds on the subtraction function remain the same even if the equality sign in (\ref{eq:c2}) is replaced by the $\leq$ sign. Moreover, as we shall prove in Sec. \ref{sec:cn}, if instead of a given
$|T_1^{\Regge}(\nu^2,Q^2)|$ we use an upper estimate of it,  the constraints on the subtraction function  become weaker. This monotonicity property will be very useful for checking the stability of the method toward the variation of the input. As we will discuss in Sec. \ref{sec:cn},  a large error, of 30\%, will be adopted for the modulus of the Regge parametrization. Thus, if the Compton amplitude contains a contribution from a fixed pole, then that
contribution is included in the r.h.s. of (\ref{eq:c2}). We do
not impose any constraints on the $Q^2$-dependence of the fixed pole term, but allow
it to be arbitrary, subject only to the constraint that the contribution from the
fixed pole does not exceed 30\% of the leading terms from Pomeron and Reggeons.

Finally, since we are interested in the subtraction constant,  we consider the  condition
\be 
 T^\inel_1(0,Q^2)=S_1^\inel(Q^2),\label{eq:S1in}
\ee
where $S_1^\inel(Q^2)$ is the inelastic part of the subtraction function (\ref{eq:S}). It satisfies the low-energy theorem  \cite{GHLR2015}
\begin{align}
  S_1^\inel(0)= - \frac{\kappa^2}{4\mN^2}-\frac{\mN}{\alphaem}\,\beta_M,\label{eq:S1in0}
\end{align}
where  $\alphaem$  is the fine structure constant, $\kappa$ is the anomalous magnetic moment of the particle and $\beta_M$ its  magnetic polarizability.

As we shall see later, the conditions (\ref{eq:inel}), (\ref{eq:c1}), (\ref{eq:c2}) and (\ref{eq:S1in}) do not specify uniquely the physical amplitude. If the input conditions are compatible among themselves, one can find  a whole class of functions analytic in the $\nu^2$ complex plane cut along  $\nu^2\ge \nuth^2$, which satisfy these conditions. Obviously, the physical amplitude $T^\inel_1$ must belong to this ``admissible" class. On the other hand, if the conditions are not mutually consistent, there is no function that can satisfy  all conditions simultaneously, {\em i.e.} the admissible class is empty. As we shall show in the next section, it is possible to formulate a necessary and sufficient condition for the existence of at least one function in this class.
 
\vspace{0.5cm}

\section{Optimal consistency condition \label{sec:optimal}}
In this section  we express the information available on the amplitude in a suitable form that allows the application of standard mathematical results \cite{Duren}, and formulate an extremum problem whose solution describes in an optimal way the consistency of the input conditions\footnote{Similar methods have been applied to other physical situations in  \cite{ Caprini:1981, Caprini1, Caprini2, CGP2014}. In particular, in Refs. \cite{Caprini1, Caprini2} it allowed the derivation of model-independent constraints on the real proton Compton scattering.}. 

 We define first a so-called outer function \cite{Duren}, denoted as $C(\nu^2, Q^2)$, which is  analytic and without zeros in the $\nu^2$ plane cut along the semiaxis $\nu^2\ge \nu^2_h$ at fixed $Q^2$, and whose modulus on this cut is given by
\beq
|C(\nu^2, Q^2)|=|T_1^{\Regge}(\nu^2,Q^2)|, \quad \nu^2 \ge \nu^2_h. \label{eq:outer}
\eeq
With this definition, we write the boundary condition  (\ref{eq:c2}) in the equivalent form
\beq
\left|\frac{T^\inel_1(\nu^2,Q^2)}{C(\nu^2, Q^2)}\right| = 1, \quad \nu^2 \ge \nu^2_h.\label{eq:c2norm}
\eeq
The function $C(\nu^2, Q^2)$ is obtained by the explicit formula\footnote{An equivalent representation can be written in the canonical $z$-variable defined in (\ref{eq:z}), see \cite{Duren}.}
\begin{align}
C(\nu^2, Q^2)=\exp\left[\D\frac {\sqrt {\nu_h^2 - \nu^2}} {\pi} \int\limits^{\infty}_{\nu^2_h} d\nu'^2\frac {\ln |T_1^{\Regge}(\nu'^2, Q^2)|}{\sqrt {\nu'^2 - \nu^2_h} (\nu'^2 -\nu^2)} \right].\label{eq:C}
\end{align}

We consider now the change of variable
 \beq\label{eq:z}
\tilde z(\nu^2)=\frac{1-\sqrt{1-\nu^2/\nu_h^2}}{1+\sqrt{1-\nu^2/\nu_h^2}},
\eeq 
which maps the $\nu^2$ plane  cut along the semiaxis $\nu^2 \geq \nu^2_h $ onto the interior of the unit disk $|z|<1$  cut along the segment $z_\thr\le x\le 1$ of the real axis, where $z\equiv \tilde z(\nu^2)$ and
\beq\label{eq:zth}
z_\thr=\tilde z(\nuth^2).
\eeq
We note also that $\tilde z(0)=0$ and the upper (lower) edge of the cut along  $\nu^2\ge \nu_h^2$  becomes the upper (lower) semicircle of the circle $|z|=1$.

 It is convenient  to define the new function
\begin{align}\label{eq:Fsigmaz}
 F(z, Q^2) \equiv \frac{  T_1^\inel(\tilde{\nu^2}(z), Q^2)  }{C(\tilde{\nu^2}(z), Q^2)}, \end{align}
 where $\tilde{\nu^2}(z)$, defined as 
\beq\label{eq:tildenu}
\tilde{\nu^2}(z)=\frac{4 z}{(1+z)^2}\,\nu_h^2,
\eeq
is the inverse of the function $\tilde z(\nu^2)$ from (\ref{eq:z}).
Since by construction $C(\nu^2, Q^2)$ is analytic and has no zeros in the  $\nu^2$-plane cut along  the semiaxis $\nu^2\ge \nu^2_h$, which in the $z$-variable  corresponds to the unit disk $\vert z\vert <1$,  the function $F(z, Q^2)$  is, for each fixed $Q^2$, a real analytic function in the unit disk $\vert z\vert <1$ except for a cut along the real segment $z_\thr\le x\le 1$.  The  discontinuity across the cut,  $\text{disc}\, F(x, Q^2)=  F(x+i\epsilon, Q^2)-  F(x-i\epsilon, Q^2)= 2\ i\ \Im F(x+i\epsilon, Q^2)$, is related to the imaginary part 
\beq\label{eq:discF}
\Im F(x+i\epsilon, Q^2)\equiv \sigma(x, Q^2), \quad z_\thr \le x \le 1, 
\eeq
which is obtained as
\beq\label{eq:sigmax}
 \sigma(x, Q^2) = \frac{\pi V_1(\sqrt{\tilde{\nu^2}(x)}, Q^2) }{C(\tilde{\nu^2}(x), Q^2)}.
\eeq
In writing this expression we took into account the fact that the  function $C(\tilde{\nu^2}(z), Q^2)$ is real and has no discontinuity across the real interval  $z_\thr\le x\le 1$.

Recalling that $\tilde z(0)=0$, we note  that $F(z, Q^2) $ satisfies also the condition
\beq\label{eq:F0}
F(0, Q^2) =S(Q^2)
\eeq
where
\beq\label{eq:Sz}
S(Q^2) \equiv \frac{S_1^\inel(Q^2)}{C(0, Q^2)}.
\eeq
The constraints (\ref{eq:discF}) and  (\ref{eq:F0}) are fully implemented by the representation\footnote{The transition point $z=1$ requires a special treatment in the intermediate steps, see. Refs. \cite{Caprini1, CGP2014}. We omit it here, since it does not affect the final result quoted in the Appendix.}
\begin{align}\label{eq:ga}
F(z, Q^2 )=S(Q^2) +\frac{z}{\pi}\int_{z_\thr}^{1}dx\,\frac{\sigma(x, Q^2)}{x(x-z)} + z  g(z, Q^2  )\,,
\end{align}
where $g(z, Q^2  )$ is, for every fixed $Q^2\ge 0$, an arbitrary function analytic in $|z|<1$, free of any constraints at interior points. The presence of this function shows that the low-energy conditions do not specify uniquely the amplitude.

 We exploit now the boundary condition  (\ref{eq:c2norm}), which  becomes, in the $z$-variable,
\beq
|F(\zeta, Q^2)|= 1, \quad |\zeta|=1.\label{eq:c2z}
\eeq
By inserting in this condition the general representation (\ref{eq:ga})    and dividing both terms of the equality (\ref{eq:c2z}) by the factor $|\zeta|\equiv 1$, we obtain with no loss of information the relation
\beq\label{eq:dif}
 |g(\zeta, Q^2) -h(\zeta, Q^2)| = 1, \quad |\zeta|=1,
\eeq
where 
\begin{align}\label{eq:h}
h(\zeta, Q^2)= -\frac{S(Q^2)}{\zeta} -\frac{1}{\pi}\int_{z_\thr}^{1}dx\,\frac{\sigma(x,Q^2 )}{x(x-\zeta)}
\end{align}
is a calculable complex function defined in terms of the input information,  and $g(z, Q^2)$ is, at fixed $Q^2\ge 0$, an arbitrary function analytic in $\vert z \vert<1$. In general, we expect that the boundary condition (\ref{eq:dif}) for a given input $h$ is satisfied by many analytic functions  $g$.  On the other hand,  it is possible that the input $h$ is such that no analytic function satisfying this condition exists.   So, the question is to decide whether the class of the functions  $g$  satisfying the condition (\ref{eq:dif}) contains at least one element. In order to answer this question, we consider the following functional extremal problem: find
\begin{equation}\label{eq:inf}
\mu_0(Q^2)=\min_{g\in H^\infty} \Vert g-h\Vert_{L^\infty}\,,
\end{equation}
 where the $L^\infty$ norm is defined as \cite{Duren}
\begin{equation}\label{Hinf}
\Vert F\Vert_{L^\infty}\equiv \sup_{\theta\in (0,2\pi)}\vert F(e^{i\theta}) \vert.
\end{equation} 
At each fixed $Q^2$,  the minimization in (\ref{eq:inf}) is done with respect to the functions $g(z, Q^2)$
analytic  in the disk $\vert z \vert<1$ and bounded on its boundary (this class of functions is  denoted as $H^\infty$ \cite{Duren}).

For what follows it is important to note that, if at least one function $g$ that satisfies the condition (\ref{eq:dif}) exists, the minimal norm $\mu_0(Q^2)$  satisfies the condition
\beq
\mu_0(Q^2)\le 1.\label{eq:optimal}
\eeq  
On the other hand, if the minimal norm $\mu_0(Q^2)$ is strictly greater than 1, the difference in the left-hand side of (\ref{eq:dif}) will be strictly greater than 1 for every analytic function $g$, which means that there are no analytic functions which can satisfy  the constraint (\ref{eq:dif}). Therefore, the inequality (\ref{eq:optimal}) is a necessary and sufficient condition for the existence of at least one admissible function $g$ that satisfies the input constraints imposed on the physical amplitude.
Implicitly, it represents a consistency condition  of the input quantities entering the function $h$, from which we can derive correlations and bounds on various input quantities.

 From the definition of the minimal norm, one can see that the problem is not changed if the equal sign in the condition (\ref{eq:c2}) is replaced by the $\leq$ sign. In the next section we shall show also that by using instead of the right-hand side of (\ref{eq:c2}) an upper estimate of the Regge modulus, we obtain, for a fixed subtraction constant, a lower value of $\mu_0(Q^2)$, which means that the allowed range for the subtraction function derived from the condition (\ref{eq:optimal}) will be larger.

 The solution of the problem (\ref{eq:inf}) and the algorithm for calculating the quantity $\mu_0(Q^2)$ are presented in the Appendix. In the next section we discuss the numerical calculation of the quantities entering the solution, based on the experimental input on the cross sections and the Regge parametrization.

\vspace{0.5cm}
\section{Numerical analysis of the input\label{sec:cn}}
As shown in (\ref{delta0}), the parameter $\mu_0(Q^2)$ is given by  the square root of the greatest
 eigenvalue of the positive-semidefinite matrix ${\cal H}^\dagger {\cal H}$, where ${\cal H}$ is a Hankel matrix defined in (\ref{hank}) in terms of the Fourier coefficients $c_k(Q^2)$ defined in (\ref{four}) or (\ref{four1}). 

Having in view the expression (\ref{eq:h}) of the function $h(\zeta, Q^2)$, it is convenient to use for the calculation of $c_k(Q^2)$   the  definition  (\ref{four1}) as a contour integral. By  applying the residue theorem, we obtain the convenient expression
\begin{align}\label{eq:cn}
c_k(Q^2) =-S(Q^2) \delta_{k,1}+ \frac{1}{\pi}\int_{z_\thr}^{1} x^{k-2} \sigma(x, Q^2) dx,~~ k\ge 1,
\end{align}
where $S(Q^2)$, $z_\thr$ and  $\sigma(x, Q^2)$ are defined in Eqs. (\ref{eq:Sz}),  (\ref{eq:zth}) and  (\ref{eq:sigmax}), respectively. The cross sections at low and intermediate energies enter through the function $V_1(\nu, Q^2)$, and the Regge model is included in the outer function  $C(\nu^2, Q^2)$ defined in (\ref{eq:C}).

For the experimental input we rely on the detailed analysis performed in the recent paper \cite{GHLR2015}. At low energies, for $W=\sqrt{s}\leq 1.3 \gev$,    we use the data on the cross sections measured in \cite{Drechsel}-\cite{chiralMAID}.  More precisely, we use as central values the average of MAID \cite{MAID} and DMT \cite{DMT} data, and ascribe to them an error equal to their difference. 
In the range $1.3<W<3 \gev$ we use the parametrization of Bosted and Christy \cite{Bosted1, Bosted2}. For the proton amplitude, the error was obtained by attaching an uncertainty of 8\% to the proton  transverse and longitudinal cross sections and adding the errors up in quadrature. For the proton-neutron difference an uncertainty of 30\% was assumed (to allow the comparison with the results reported in \cite{GHLR2015}, for $Q^2\ge 0.5 \gev^2$ the error was assumed to be  8\%  from the proton cross section, calculated as mentioned above). 

 Several  Regge parametrizations of the cross sections at high energies have been proposed in the literature \cite{GVMD}-\cite{Sibirtsev}. We shall use at $W\geq W_h$, where $W_h=3\, \gev$, the  parametrizations given by the vector-meson dominance model of Alwall and Ingelman  \cite{GVMD}: 
\begin{align}
 \sigma_{\indT} &= \beta^{\indT}_{\indP}(Q^2)s^{\alpha_P-1}+\beta_R^{\indT}(Q^2) s^{\alpha_R-1}\co\no
\sigma_{\indL}&= \beta^{\indL}_P(Q^2)s^{\alpha_P-1}+\beta_R^{\indL}(Q^2) s^{\alpha_R-1}\co
\label{eq:AI} 
\end{align}
where $s=W^2$ is the square of the center-of-mass energy. A ``soft" Pomeron with intercept  $\alpha_P=1.091$ is adopted,  while  the Reggeons with the quantum numbers of $f$ and $a_2$ are represented by a single contribution with $\alpha_R= 0.55$. 
The Pomeron residues of proton and neutron are usually assumed to be the same:
\be
\label{eq:betaP} 
\beta^{\indT}_P(Q^2)^n= \beta^{\indT}_P(Q^2)^p\co\quad
\beta^{\indL}_P(Q^2)^n= \beta^{\indL}_P(Q^2)^p\fs
\ee
while for the Reggeons one uses \cite{Pilkuhn:1973wq}:
\begin{align}
 \beta_R^{\indT}(Q^2)^n=\xi \,\beta_R^{\indT}(Q^2)^p\co\quad
 \beta_R^{\indL}(Q^2)^n=\xi\, \beta_R^{\indL}(Q^2)^p\co
\label{eq:betar}
\end{align}
with $\xi \simeq 0.74$. 

\begin{figure}[ht]\vspace{0.7cm}
\centering
\includegraphics[width=\linewidth, width=7cm]{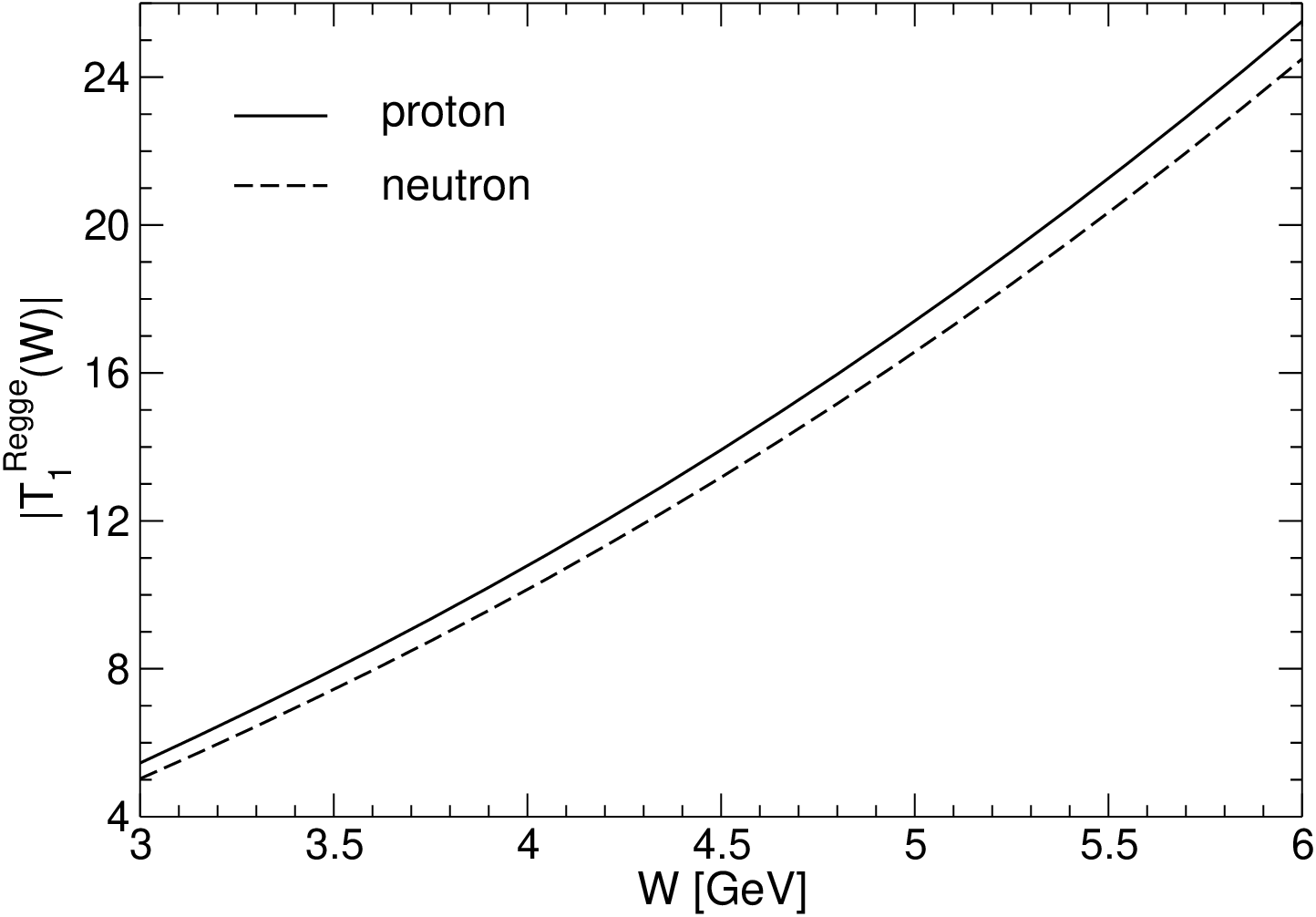} 
\caption{Modulus of the Regge parametrization of the amplitude $T_1^{\Regge}$ for $Q^2=0$, in the region $3\leq W\leq 6 \gev$. 
\label{fig:Regge} \vspace{0.4cm} }
\end{figure}

 The structure function $V_1(\nu, Q^2)$ at $\nu < \nu_h$ is obtained directly from the cross sections using (\ref{eq:VsigmaLT}).  At higher energies, at $\nu \ge \nu_h$, we can  recover locally  the full amplitude $T_1^\inel$ from  the parametrization  of the imaginary part obtained from (\ref{eq:AI}). Using the standard Regge expression (\ref{eq:T1RStandard}), it is easy to see that this is achieved by making in (\ref{eq:AI}) the replacements
\begin{align}\label{eq:recover}
 \beta_j \rightarrow - \frac{1+\exp[-i\pi\alpha_j]} {\sin\pi \alpha_j }\, \beta_j, \quad j=P,R.
\end{align}
By this prescription we ensure that the real and imaginary parts are related as required by the Regge model: for the Pomeron, the real part is much smaller than the imaginary one, while for the Reggeons, using $\alpha_R=0.55$, the real and imaginary parts are almost equal. 

In Fig. \ref{fig:Regge} we show for illustration the modulus of the amplitude $T_1^{\Regge}$ obtained with this prescription for  $Q^2=0$, in the region $3\leq W\leq 6 \gev$. The small difference between the proton and the neutron amplitudes is due to the different Reggeons contribution and decreases at higher energies, where the Pomeron dominates (for instance, the difference is 8\% at 3 GeV and only  2 per mille at 100 GeV).  We emphasize that the expression obtained  by the prescription (\ref{eq:recover}) is used only on the unitarity cut at fixed $Q^2$ for $\nu^2\ge \nu_h^2$, where $\nu_h=(W_h^2-m^2+Q^2)/2m$ with $W_h=3\, \gev$. No assumptions on the analytic continuation or the analytic properties of the Regge parametrization at other points of the $\nu^2$-complex plane are made.  As already mentioned, an error of 30\%  was attached to the modulus of the Regge parametrization in our analysis.

Since we use as input different models at
various energies, one might ask what are the consequences of matching these models for the accuracy of the numerical analysis.  Actually, the large uncertainties that we adopted cover to some extent the discrepancies at the matching points. It is important to emphasize also that, in contrast to the usual FESR, the present dispersive formalism does not involve the calculation of principal values or other quantities sensitive to the discontinuities of the input on the cut. Therefore, the details of the matchings have a smaller effect in our approach than in the standard dispersion relations. 

Using the expressions  (\ref{eq:cn}) and (\ref{delta0}) of the Fourier coefficients and the minimal norm $\mu_0$, it is easy to prove  a monotonicity property already mentioned above. Assume that we replace the input modulus at fixed $Q^2$ given in the right-hand side of  Eq. (\ref{eq:c2}) by an estimate of it from above. Then the real and positive values of the outer function $C(\nu^2, Q^2)$ defined in (\ref{eq:C}) will increase at all  real points $\nu^2$ below $\nu_h^2$. From the definitions (\ref{eq:sigmax}) and (\ref{eq:Sz})  it follows that the real quantities $\sigma(x, Q^2)$ and $S(Q^2)$  entering the expression (\ref{eq:cn}) will decrease if  $S_1^\inel(Q^2)$ is kept fixed. This will lead to smaller values for all the coefficients $c_k$ and therefore to a smaller norm $\mu_0$. As a consequence, the allowed range of the parameter  $S_1^\inel(Q^2)$, derived from the consistency condition (\ref{eq:optimal}), will be larger. 

This property has important consequences. First, it allows us to use  at all energies above $W=3 \gev$  the parametrization of the Pomeron by a ``soft" trajectory with an intercept $\alpha_P$ slightly larger than 1. This model  is valid strictly speaking only at finite energies, since it predicts an asymptotic increase in conflict with unitarity. Indeed,  universality implies that the same Pomeron would also show up in hadronic scattering processes and
 that would violate the Froissart bound (an attempt at describing deep inelastic scattering at high energies with a
logarithmic high-energy behavior consistent with the Froissart bound is presented in  \cite{Block}).
Actually, the soft Pomeron is expected to overestimate the physical logarithmic increase with energy of the amplitude. Since the bounds are monotonically decreasing functionals of the input modulus,  the soft Pomeron is expected to lead to weaker bounds on the subtraction functions compared to those obtained with the true behavior, preserving the validity of the method.  

In view of the same monotonicity property, the weakest upper and lower bounds are obtained with the modulus increased within the error channel. The large uncertainty of 30\% which we adopted covers possible variations of the subasymptotic terms related to the standard Regge poles and also a possible real constant, which can arise due to a fixed pole at $J=0$ in the angular momentum plane. 

  As shown in (\ref{eq:S1in0}),  the subtraction function at $Q^2=0$ is determined by the magnetic polarizability.  The experimental values quoted in the recent papers \cite{McGovern:2012ew} and \cite{Myers:2014ace}, expressed  in units of $10^{-4}\ \text{fm}^3$, are\footnote{Note also the average values given in PDG \cite{PDG}, $\beta_M^p = 2.5\pm 0.4$ and $\beta_M^n = 3.7\pm 1.2$.} 
\beq\label{eq:beta}
\beta_M^p = 3.15\pm 0.50, \quad \beta_M^n = 3.65\pm 1.50, 
\eeq
which imply $\beta_M^{p-n}=-0.5 \pm 1.6$.  Using these values in the expression (\ref{eq:S1in0}), we obtain the subtraction functions
\begin{align}\label{eq:S0}
S_{1,p}^\inel(0)&=(-6.18 \pm 0.84)\gev^{-2}\co \no
S_{1,n}^\inel(0)&=(-7.15 \pm 2.51)  \gev^{-2}\co \no  
S_{1,p-n}^\inel(0)&=(0.96\pm 2.68)  \gev^{-2}.
\end{align} 

 As discussed above, the method requires  the calculation of the Fourier  coefficients $c_k$ given in (\ref{eq:cn}) and of the quantity $\mu_0$ given by the norm  (\ref{delta0}) of the Hankel matrix (\ref{hank}). The numerical algorithm implies actually the truncation of the Hankel matrix at a finite order and the calculation of its norm with standard programs.  In practice, we obtained good stability with the first 20-30 coefficients.
\vspace{0.5cm}
\section{Results\label{sec:applic}}
 As discussed above, the inequality (\ref{eq:optimal}) represents a necessary and sufficient condition for the consistency  with analyticity of the amplitude $T_1^\inel(\nu^2, Q^2)$. One can exploit it either for testing the consistency of the various pieces of the phenomenological input, or for deriving bounds on the subtraction function $S_1^\inel(Q^2)$.

\subsection{Consistency of the input at $Q^2=0$\label{sec:comp}}
At $Q^2=0$, where the subtraction functions are known in terms of polarizabilities as shown in (\ref{eq:S0}), we have all the ingredients to calculate the coefficients $c_k$ defined in (\ref{eq:cn}) and the parameter $\mu_0$ defined in (\ref{delta0}). Keeping fixed the subtraction functions at the central values (\ref{eq:S0}), we obtain
\begin{align}\label{eq:delta0p+n}
\mu_0^p&=  1.099 \pm 0.054_{\text{MAID}}\pm 0.053_{\text{BC}}\pm 0.099_R, \no
\mu_0^n&=  1.066 \pm 0.053_{\text{MAID}}\pm  0.053_{\text{BC}}\pm 0.254_R, \no
\mu_0^{p-n}&=  1.407  \pm 0.633_{\text{MAID}} \pm  0.199_{\text{BC}} \pm 0.325_R, 
\end{align}
where we indicated the effect of the separate variation of the MAID and DMT data \cite{MAID,DMT}, Bosted and Christy parametrization \cite{Bosted1, Bosted2} and the Regge parametrization 
\cite{GVMD}, within the uncertainties specified in the previous section. Adding in quadrature these effects leads to
\begin{align}\label{eq:delta0p+n1}
\mu_0^p& = 1.099 \pm 0.265\co\no
\mu_0^n&=  1.066 \pm 0.140\co\no
\mu_0^{p-n}&=  1.407 \pm 0.739. 
\end{align}
We recall that the consistency of the input requires values of $\mu_0$  less than 1. From (\ref{eq:delta0p+n1}) one can see that, for the  proton and neutron amplitudes, this condition is slightly violated by the central values of the input quantities, but consistency is achieved when the errors are taken into account.  We note that, for fixed values of the subtraction functions, in all cases  $\mu_0$ decreased if the low energy cross sections or the Regge modulus were increased within errors.

The central value of $\mu_0^{p-n}$  is significantly larger than 1, indicating that the central values of the various parts of the input are inconsistent with analyticity. This is due to the fact that  the   amplitude is obtained as the difference of the proton and the poorly known neutron amplitude, and is not very reliable.  Nevertheless, consistency with analyticity is achieved also in this case when the large errors discussed in the previous section are taken into account.

In the above calculations the subtraction functions $S_1^\inel(0)$ have been kept fixed at the central values given in (\ref{eq:S0}). One may reverse the argument and calculate $\mu_0$ as a function of this parameter,  keeping the cross sections and the Regge parameters fixed. We shall apply this procedure in the next subsection, where we shall derive upper and lower bounds on the subtraction function $S_1^\inel(Q^2)$ in terms of the cross sections and Regge parameters.

We end this subsection by noting that the formalism considered in the present paper can be used in principle also for testing the presence or the absence of fixed poles in the scattering amplitude. This subject was put forward a long time ago  \cite{Creutz, DaGi} and received attention also recently \cite{Brodsky:2008qu,  Gorc1, Muller:2015vha}. As is known, a fixed pole at $J=0$ in the angular momentum plane  contributes with  a real constant to the high-energy behavior of the amplitude. In the present formalism, the need of an additional term in the high-energy behavior is signaled by  values of the minimum norm $\mu_0$  larger than 1 when calculated with only standard Regge poles in the condition (\ref{eq:c2}). The results  given above show that, at $Q^2=0$, this does not seem to be the case: the analyticity test is passed within the relatively large uncertainties of the present data.  If more precise data in the future will lead to  $\mu_0>1$ even when the input will be varied within errors, it will be possible to conclude that  a fixed pole is necessary and, by adjusting the magnitude of the additional constant until the value of the quantity  $\mu_0$ becomes smaller than unity, to find constraints on its residue.

\subsection{Bounds on the subtraction function $S_1^\inel(Q^2)$ }\label{sec:Q2gt0}

 Using  arguments presented in \cite{Caprini1, CGP2014},  one can show that, for given input values of cross sections and Regge parametrizations, the minimum norm $\mu_0$ is  a convex function of the parameter $S_1^\inel(Q^2)$, displaying a single minimum.  Therefore,  from the inequality (\ref{eq:optimal}) one can derive exact upper and lower bounds on the subtraction function $S_1^\inel(Q^2)$.  The bounds define  an allowed interval for this quantity. Of course, if the values of  $\mu_0$ stay above unity for all values of the parameter  $S_1^\inel(Q^2)$,  we conclude that there are no analytic functions satisfying the  input conditions, therefore the input data are inconsistent with analyticity.

\begin{figure}[ht]
\centering \vspace{0.4cm}
\includegraphics[width=\linewidth, width=7cm]{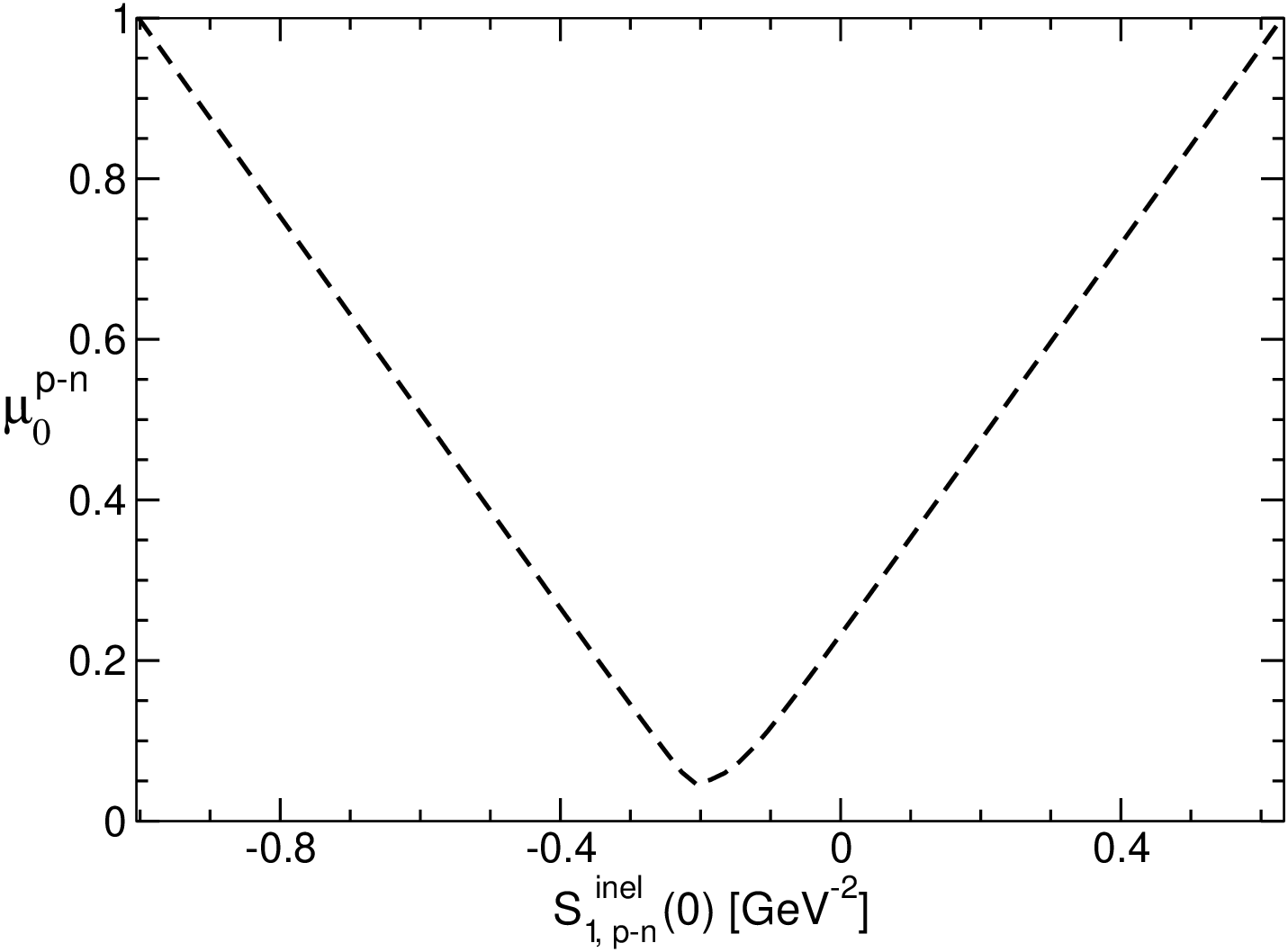} 
\caption{Dependence of $\mu_0^{p-n}$ on the subtraction constant  $S_{1, p-n}^\inel(0)$.
\label{fig:1} \vspace{0.4cm} }
\end{figure}

We illustrate first the above arguments for the amplitude corresponding to the proton-neutron difference at $Q^2=0$. In Fig. \ref{fig:1}, we present $\mu_0^{p-n}$ as a function of the subtraction function $S_{1, p-n}^{~\inel}(0)$, showing for convenience only the range where the consistency condition (\ref{eq:optimal}) is satisfied. From this condition, we obtain the allowed range $-1.00\leq S_{1, p-n}^\inel(0)\leq 0.64 \gev^{-2}$. By  taking into account the uncertainties of the cross sections and the Regge parametrization,  this range becomes
\beq\label{eq:S1pnbounds}
-1.59\leq S_{1, p-n}^\inel(0)\leq 1.22 \gev^{-2}.
\eeq
In this calculation we varied independently the MAID/DMT cross sections below 1.3 GeV, the parametrization of Bosted and Christy  in the range $1.3<W<3 \gev$ and the modulus at higher energies, and modified the bounds by the quadratic sum of the separate shifts.  We note that the central experimental  value $S_{1, p-n}^\inel(0)=0.96\gev^{-2}$ given in (\ref{eq:S0}) is slightly above the upper bound $0.64 \gev^{-2}$ obtained with the central input on the unitarity cut,   which was reflected by the central value $\mu_0^{p-n}>1$ given in  (\ref{eq:delta0p+n1}). However, again in accordance with (\ref{eq:delta0p+n1}), the central experimental  value is within the enlarged  range  (\ref{eq:S1pnbounds})  which includes the uncertainties of the input. We emphasize that this range  is  narrower than the experimental interval $-1.71 \leq S_{1, p-n}^\inel(0)\leq 3.64 \gev^{-2}$ which follows from (\ref{eq:S0}), and is consistent with the value $S_{1, p-n}^\inel(0)=-0.3(1.2) \gev^{-2}$ derived in \cite{GHLR2015}.

\begin{figure}[htb]\vspace{0.3cm}
\centering
\includegraphics[width=\linewidth, width=8cm]{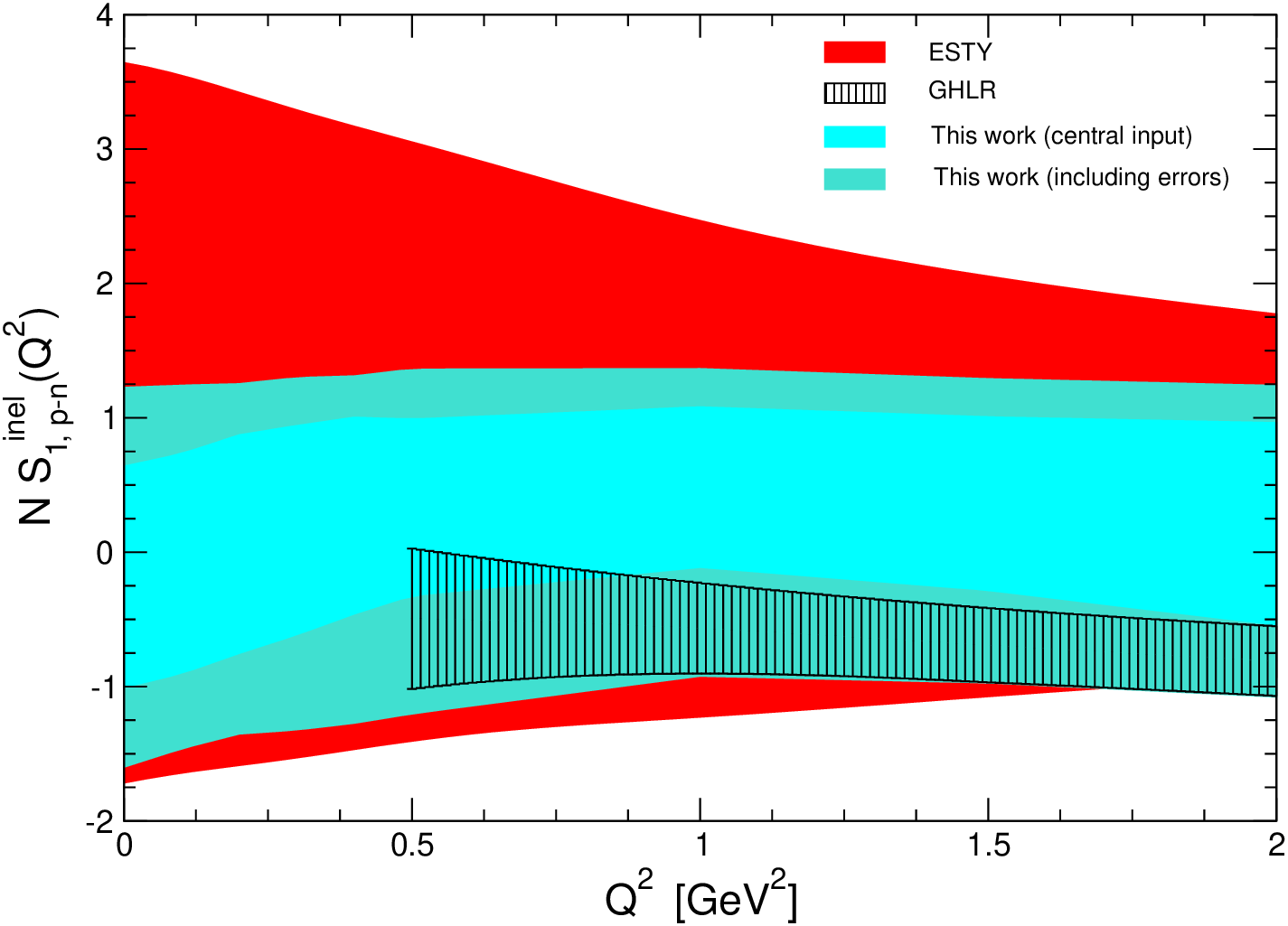} 
\caption{Bounds on  the proton-neutron subtraction function  $S_{1, p-n}^\inel(Q^2)$ for various $Q^2$. The red band is parametrization (\ref{eq:ansatzE}) proposed in \cite{Erben:2014hza}. The band denoted GHLR was derived in \cite{GHLR2015}. Cyan band: allowed range  situated between  the lower and upper bounds obtained from the condition (\ref{eq:optimal}), with the central values of the cross sections and Regge parameters. Turquoise band: enlarged range taking into account the uncertainties of the input. As in  \cite{GHLR2015},  for an easier comparison the values on the vertical axis are multiplied by the factor $N=(1+Q^2/M_d^2)^2$, with $M_d^2=0.71 \gev^2$.
\label{fig:2} \vspace{0.2cm} }
\end{figure}

\begin{figure*}[ht]
\centering
\includegraphics[width=\linewidth, width=6.5cm]{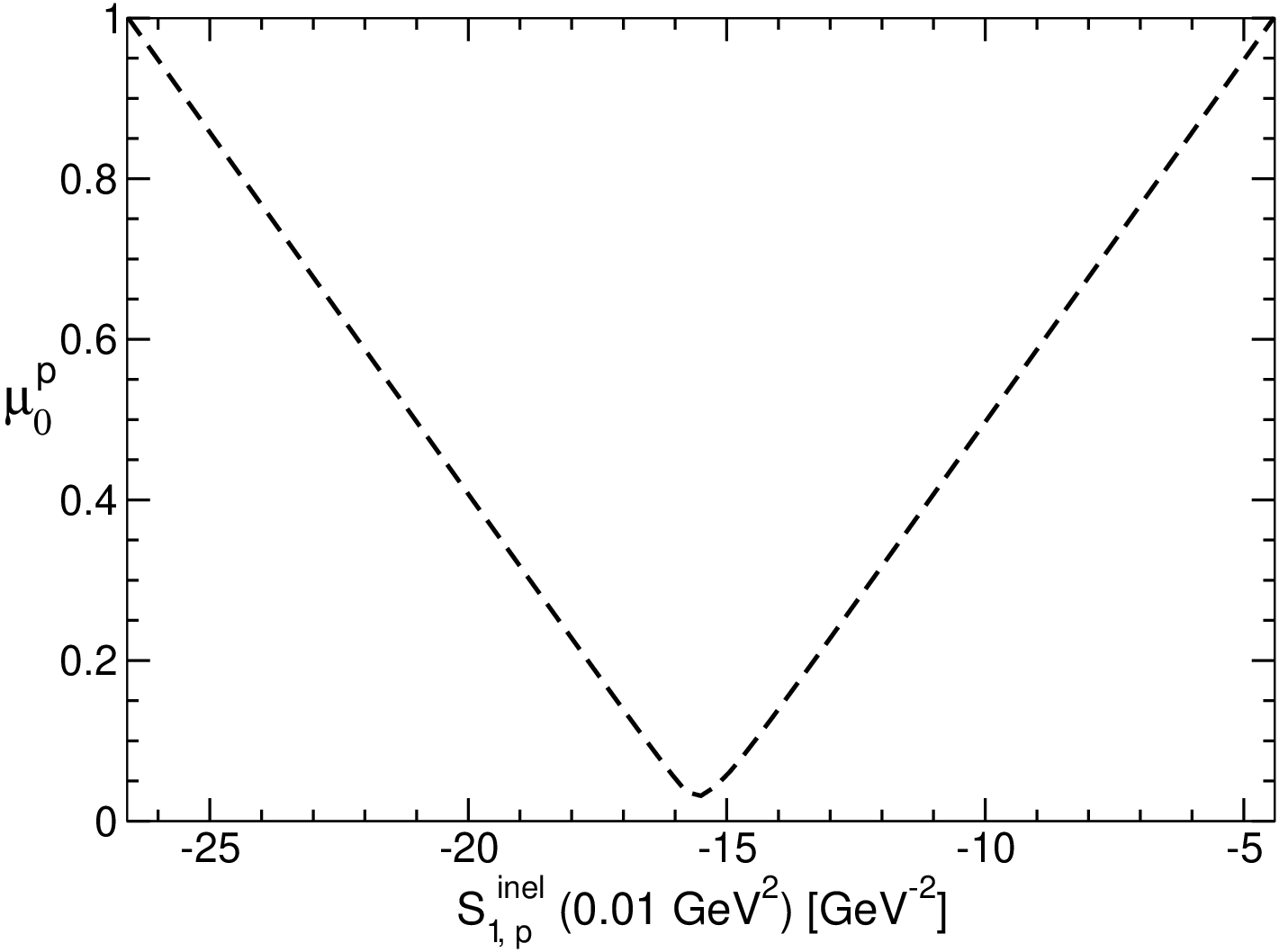} \hspace{1.4cm} 
\includegraphics[width=\linewidth, width=6.5cm]{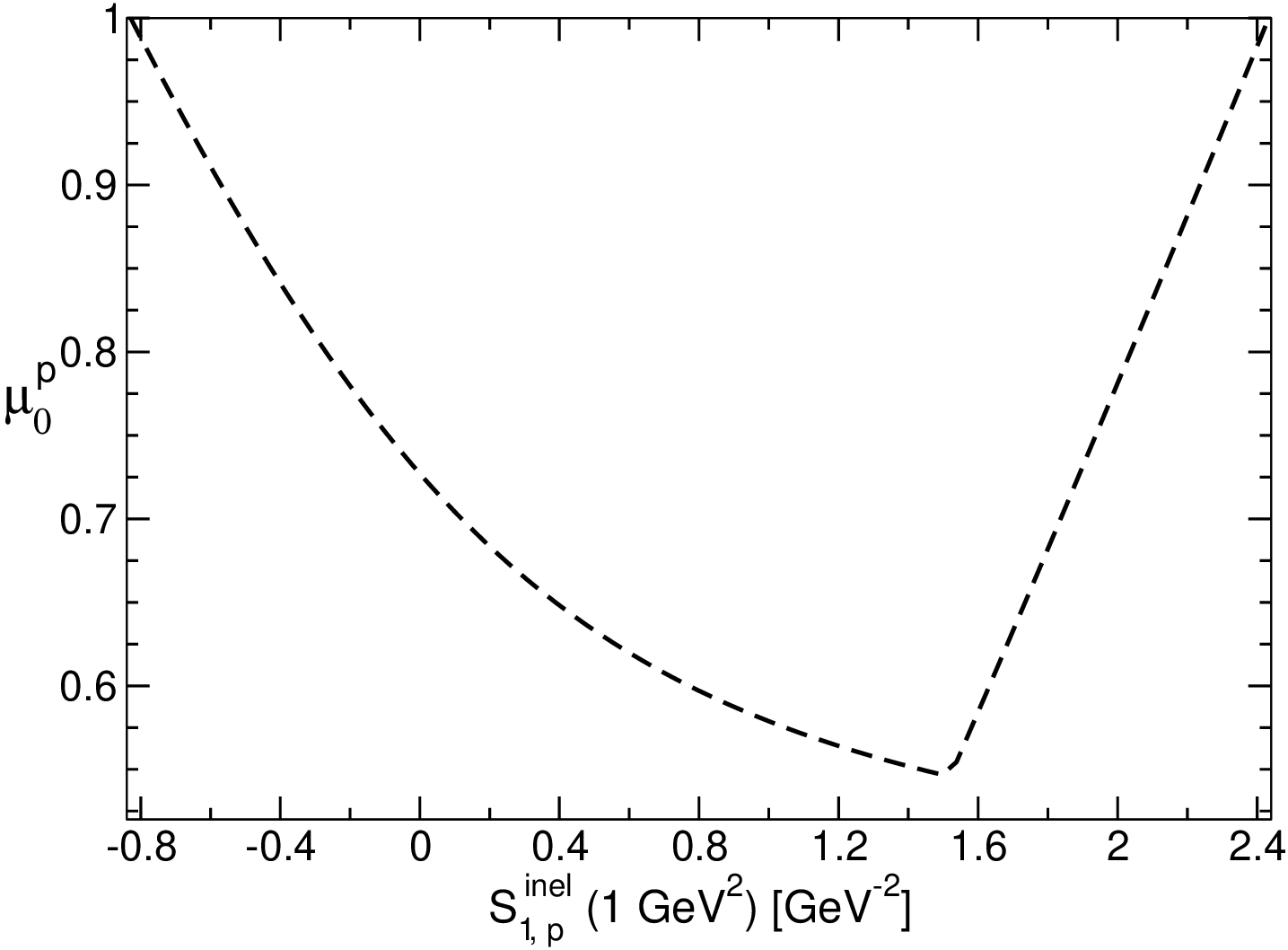}
\caption{Dependence of $\mu_0^{p}$ on the subtraction constant  $S_{1, p}^\inel(Q^2)$, for two values of $Q^2$: $Q^2=0.01 \gev^2$ (left) and  $Q^2=1 \gev^2$ (right).
\label{fig:3} \vspace{0.3cm} }
\end{figure*}

By repeating the calculations  for other values of $Q^2$, we obtain the bounds for $S_{1, p-n}^\inel(Q^2)$ presented in Fig. \ref{fig:2}, where the cyan and turquoise bands are the allowed ranges obtained with central input and including the uncertainties. In the same figure we show for comparison the allowed band (denoted as GHLR)  obtained in \cite{GHLR2015} from the assumption of ``Reggeon dominance", {\em i.e.} by requiring  the absence of constant terms in the asymptotic behavior of the amplitude.  One can see that our results are considerably weaker, which was to be expected since we work in a less restrictive formalism, where the absence of the fixed pole is not imposed in a manifest way. It is important to note however that our results are consistent within uncertainties with the allowed band derived in \cite{GHLR2015}. 

 As already mentioned, the subtraction function was assumed   to be completely independent of the cross sections in Refs. \cite{WalkerLoud:2012en,Thomas:2014dxa, Erben:2014hza}, where  rather arbitrary parametrizations of this function  have been proposed.   It is instructive to compare our results also with these parametrizations. As discussed in \cite{GHLR2015}, the expression proposed in  \cite{WalkerLoud:2012en} is not consistent with the short-distance properties of QCD. Therefore, we shall consider only  the variant adopted in \cite{Erben:2014hza}, where the inelastic part of the subtraction function reads 
\begin{align}
 S_\text{1, ESTY}^\inel(Q^2)&=-\left(\frac{m_1^2+c\,Q^2}{m_1^2+Q^2}\right)\left(\frac{m_1^2}{m_1^2+Q^2}\right)^{2}\frac{\mN\beta_M}{\alphaem}\no
 &-\frac{4\mN^2\{G_E(Q^2)-G_M(Q^2)\}^2}{(4\mN^2+Q^2)^2}.
 \label{eq:ansatzE}
\end{align} 
 We evaluated this expression using  the Kelly parametrizations  \cite{Kelly} for the nucleon form factors, the range of $\beta_M^{p-n}$ given below (\ref{eq:beta}) and the values  $c=0$ and $m_1= \sqrt{3} m_\beta$ with $m_\beta=0.46\pm 0.1\pm 0.04 \gev$ from  \cite{Erben:2014hza}. The result is represented in Fig. \ref{fig:2} as the red band denoted as Erben-Shanahan-Thomas-Young (ESTY).  One can see that the allowed band derived in the present framework is somewhat  narrower than the band obtained with the ESTY ansatz.

By performing the same analysis in the proton case, we note first that the allowed range obtained at $Q^2=0$ with central input, namely $-30.2 \leq  S_{1, p}^\inel (0) \leq -7.33  \gev^{-2}$, is quite large. This can be explained by the fact that, due to the Pomeron,  the proton amplitude is allowed to increase more rapidly on the unitarity cut. By invoking the maximum modulus principle, this implies  weaker constraints at points below the cut.  We see however that, although the interval is large, the upper bound  lies actually below the central experimental value given in (\ref{eq:S0}), obtained from the magnetic  polarizability (this discrepancy was already signaled actually by the central value of $\mu_0^p$ larger than 1  given in (\ref{eq:delta0p+n1})).  By taking into account the errors of the cross sections and the 30\% uncertainty of the Regge modulus, which is quite large in the proton case, the upper bound is pushed up to
\beq\label{eq:Sp01}
 S_{1, p}^\inel (0)\ \leq -3.77  \gev^{-2},
\eeq
becoming consistent with the experimental range (\ref{eq:S0}). This exercise shows that the upper bounds  calculated with more accurate data on the cross sections might put a nontrivial constraint on the proton magnetic polarizability.

 At nonzero values of $Q^2$, it turns out that the input is consistent in the sense explained above  only for the intervals $Q^2\leq 0.1 \gev^{2}$ and   $Q^2\geq 0.5 \gev^2$. For illustration we show in 
 Fig. \ref{fig:3} the variation of $\mu_0^{p}$ with the quantity $S_{1, p}^\inel(Q^2)$ for two values of $Q^2$, one close to 0 and the other equal to $1 \gev^2$.  Central values of the low-energy cross sections and Regge parameters have been used in the calculation. In both cases this input is consistent, in the sense that there are values of the subtraction constant for which the compatibility condition (\ref{eq:optimal}) is satisfied. For convenience, we show  only the ranges of $S_{1, p}^\inel(Q^2)$ that satisfy this  condition. From the figure, one can read the upper and lower bounds 
$-26.6 \leq  S_{1, p}^\inel (0.01 \gev^2) \leq -4.4  \gev^{-2}$ and
$ -0.84 \leq   S_{1, p}^\inel (1 \gev^2) \leq 2.44  \gev^{-2}$.
We note that the allowed ranges of $S_{1, p}^\inel (Q^2)$ for  $Q^2$ close to 0 are very large,  which shows that the constraining power of the formalism in this domain is weak.  However, as in the case $Q^2=0$ discussed above, the upper bounds might lead to nontrivial constraints. 

\begin{figure}[ht]\vspace{0.7cm}
\centering
\includegraphics[width=\linewidth, width=8cm]{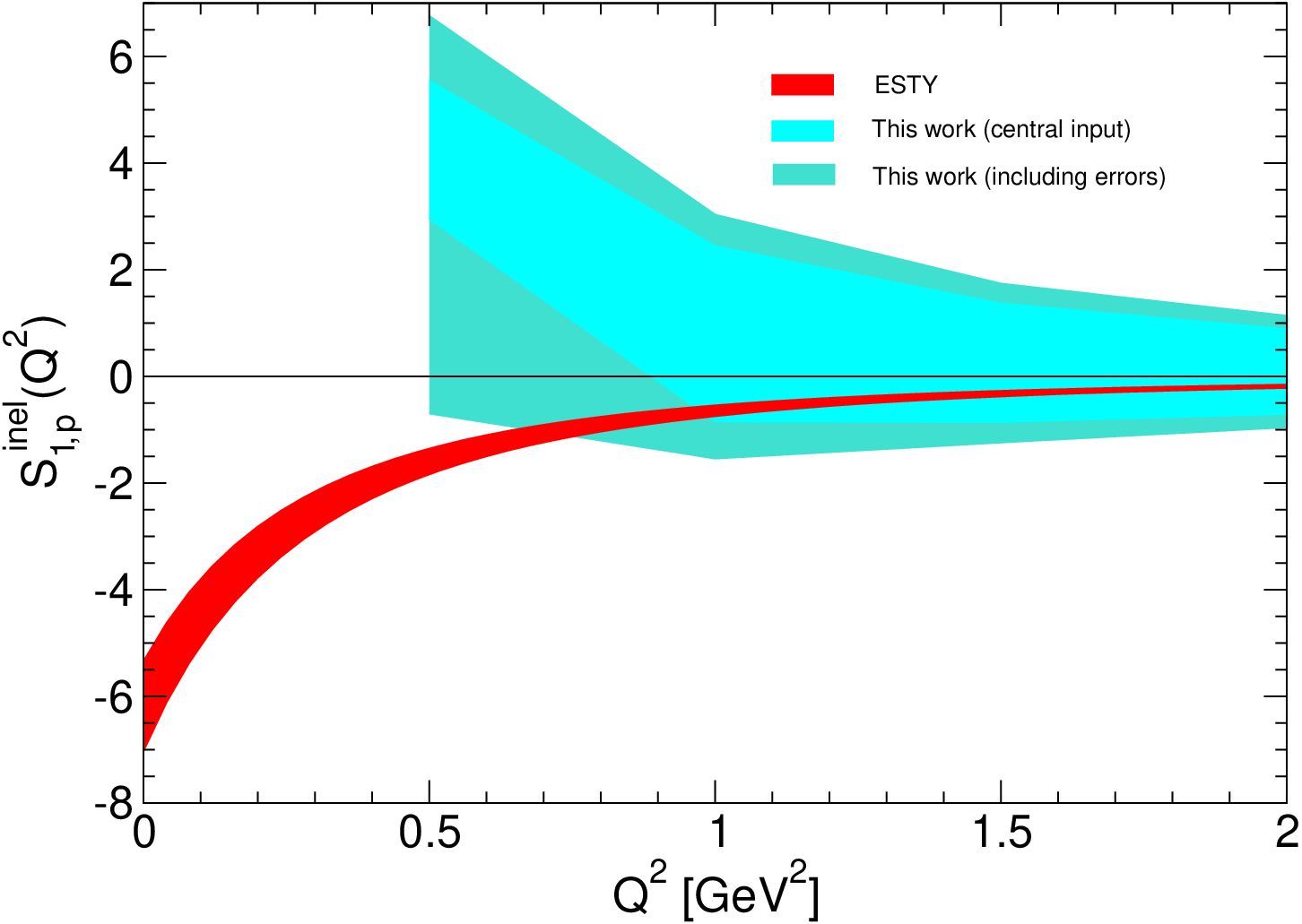} 
\caption{Allowed ranges of the proton subtraction function  $S_{1, p}^\inel(Q^2)$ for $Q^2\ge 0.5 \gev^2$, compared with the parametrization (\ref{eq:ansatzE}).
\label{fig:4} \vspace{0.4cm} }
\end{figure}
As we mentioned, for $0.1< Q^2 <0.5 \gev^2$, the parameter  $\mu_0^p$  was larger than unity for all values of the subtraction function  $S_1^\inel(Q^2)$  (more exactly, the minimum value of  $\mu_0^p$ in figures similar to Fig. \ref{fig:3} turned out to be larger than 1, and did not decrease below unity even when the various pieces of the input were varied independently within errors). Therefore, in  Fig. \ref{fig:4} we show the allowed band for the subtraction function calculated in this work only for $Q^2\ge 0.5\gev^2$. Although quite broad, the band is not consistent with the  ESTY ansatz (\ref{eq:ansatzE}) near  $Q^2 = 0.5 \gev^2$: the bounds suggest the possibility of positive values for $S_{1, p}^\inel(Q^2)$, which are not allowed by the ansatz. Of course, in view of the large uncertainties in the data and the fact that for $Q^2< 0.5\gev^2$ the structure functions and the Regge parameters are in conflict with analyticity, this result  should be taken only as a very preliminary one. On the other hand, the allowed band shown in  Fig. \ref{fig:4} is consistent, within uncertainties, with the behavior derived recently in \cite{ToVa} by a dispersive analysis, assuming the absence of $J=0$ fixed poles\footnote{I thank the authors of \cite{ToVa} for performing this comparison.}.

\section{Summary and conclusions\label{sec:summary}}
In this paper we proposed and investigated an alternative dispersive framework for the virtual Compton  scattering on the nucleon. The 
work was motivated by the recent interest in the study of this process in connection with the Cottingham formula for the proton-neutron electromagnetic mass difference, the nucleon polarizabilities and the proton radius puzzle. We considered in particular the inelastic part $T_1^\inel(\nu^2, Q^2)$ of one of the invariant amplitudes,  which requires a subtraction in the standard dispersion relation  at fixed $Q^2$. 
The aim was to  investigate in a parametrization-free approach the subtraction function  and the possible existence of  fixed poles at $J=0$ in the angular momentum plane.  Our formalism does not rely on the hypothesis that the Compton amplitude is free
of fixed poles, but covers that case as well.  So, it is less restrictive than the recent analyses \cite{GHLR2015, ToVa}, where the absence of fixed poles in the amplitude was explicitly assumed, or in the FESR studies \cite{ Gorc1}, where the contribution of a fixed $J=0$ pole was explicitly included. The price paid for this generality is the fact that our method cannot lead to definite predictions, but gives
only upper and lower bounds on the subtraction function. Within the large uncertainties of the experimental input, our results for the proton-neutron difference are consistent with those obtained in \cite{GHLR2015} and confirm the conclusions of that work.

The formalism considered in the present work uses as input the imaginary part of the amplitude obtained from electron-nucleon cross sections at low and moderate energies, and the modulus of the amplitude  at high energies, where we can recover  the real part of the amplitude from its imaginary part using the standard Regge relation (\ref{eq:T1RStandard}). 
The main mathematical result is the definition of the quantity $\mu_0$, given by the solution of the  functional minimization problem (\ref{eq:inf}), such that the inequality (\ref{eq:optimal}) is a necessary and sufficient condition for the consistency of the phenomenological input with analyticity and unitarity.  The parameter $\mu_0$ can be computed by a simple numerical algorithm as the norm of a Hankel matrix (\ref{hank}) constructed from the Fourier coefficients $c_k$, calculable according to  (\ref{eq:cn}) in terms of the cross sections and the Regge parameters. For a given phenomenological input on the unitarity cut, the quantity $\mu_0$ is a convex function of the unknown subtraction function. Then, the inequality  (\ref{eq:optimal}) allows us to find rigorous upper and lower bounds on the subtraction function in terms of this input.

 As discussed in Sec. \ref{sec:cn}, we assumed large uncertainties for the input cross sections and the Regge model. In particular, the 30\% uncertainty adopted for the modulus of $T_1^\inel(\nu^2, Q^2)$ above 3 GeV covers possible variations of the subasymptotic terms related to the Pomeron and the Reggeons and can accommodate also constant terms of a certain magnitude in the asymptotic behavior.  As proved in Sec. \ref{sec:cn},  the bounds are monotonically decreasing functionals of the input modulus. This important property allowed us to use the phenomenological parametrization by the soft Pomeron  adopted from \cite{GVMD} also at higher energies, where it overestimates the logarithmic asymptotic increase of the amplitude required by unitarity. Therefore, the bounds derived in the present work are weaker than what would be obtained with the physical asymptotic increase, which means that our approach is a conservative one.

 In Sec. \ref{sec:applic} we applied first the formalism for testing the consistency  with analyticity of the available phenomenological input on the amplitude $T_1^\inel(\nu^2, Q^2)$ at $Q^2=0$.  The central values of the electroproduction cross sections,  the Regge parametrization and the magnetic polarizabilities lead to values of $\mu_0$ which slightly violate the condition (\ref{eq:optimal}), but consistency is achieved when uncertainties of the cross sections and Regge parametrization are taken into account. 

The formalism can be applied also for investigating the possible existence of  fixed poles at $J=0$ in the angular momentum plane, which contribute with a real constant to the high-energy behavior of the amplitude. The presence of a fixed pole is revealed in principle by  a clear inconsistency of the low-energy input with the standard Regge model, signaled by large values of the parameter $\mu_0$.  Of course,  in practice inconsistency alone  would not
imply the existence of a fixed pole, it could simply mean that
uncertainties in the rest of the input (cross sections, subasymptotic
terms in the Regge amplitude) are larger than expected.  Therefore, more precise and reliable data are necessary in order to establish, from the consistency with analyticity of the input, that a fixed pole of a certain residue is required or not. 

Finally, we applied the formalism for deriving constraints on the subtraction function $S_1^\inel(Q^2)$ at various values of $Q^2$, by exploiting the fact that the quantity $\mu_0$  is a convex function of the value of this parameter.  We performed this calculation for the  proton-neutron difference and the proton amplitudes.

For the proton-neutron difference, the allowed range of $S_{1, p-n}^\inel(0)$ derived by the present method is slightly smaller than the experimental range (\ref{eq:S0}) derived from polarizabilities.   For  $Q^2\geq 0.5\gev^2$ our results are consistent within errors with the band derived  in \cite{GHLR2015} by imposing explicitly the absence of the fixed poles. Such an assumption is not explicitly adopted in our formalism, which explains the weaker results that we obtain.

 For the proton amplitude, we found that for $Q^2$ between 0.1 and $0.5\gev^2$ the phenomenological cross sections and the Regge input are mutually inconsistent, leading to values of the quantity $\mu_0$  larger than 1. For $Q^2$ near 0, the bounds derived  are quite weak, however at  $Q^2=0$  the upper bound on  $S_{1, p}^\inel(0)$ calculated with central input is in slight conflict with the experimental value obtained from the magnetic polarizability. This suggests that with  more precise data the formalism might lead to nontrivial predictions. For $Q^2\ge 0.5\gev^2$, the bounds derived in the present work are in agreement within uncertainties with the behavior found in \cite{ToVa} by assuming the absence of $J=0$ fixed poles.

To summarize, the main benefit of the present formalism is that it allows us to make predictions on the subtraction function in virtual Compton scattering with no explicit assumption about the existence or the absence of a $J=0$ fixed pole.  Thus, we obtain model-independent constraints on the subtraction function, which can be calculated in standard dispersion relations  only by adopting a specific assumption about the fixed poles.  
The numerical results reported in this paper depend on the phenomenological input available at present, which has large uncertainties. Therefore, they should be taken only as very preliminary results.    Our aim was mainly to emphasize the usefulness of a modified dispersion approach for the study of virtual Compton scattering. 

\subsection*{Acknowledgments} 
I would like to thank  J. Gasser and H. Leutwyler for  very useful discussions and detailed information concerning the structure functions used in the analysis. This work was supported by  UEFISCDI under Contract Idei-PCE No 121/2011.

\appendix*
\section{Solution of the extremal problem}

We shall find the solution of the minimization problem (\ref{eq:inf}) by
 applying a ``duality theorem" in functional  optimization theory
\cite{Duren}, which  replaces the original minimization problem (\ref{eq:inf})
by a maximization problem on a different functional space, denoted as the
``dual" space. The new problem will turn out to be easier than the original one, 
allowing us to obtain the solution by means of a numerical algorithm. 

Using standard terminology \cite{Duren}, we  denote by $H^p$, $p<\infty$,  the class of functions $F(z)$ which are
analytic inside the unit disk $|z|<1$ and satisfy the boundary condition
\begin{equation}\label{Hp}
\|F\|_{L^p}\equiv \left[\frac{1}{ 2\pi}\int_0^{2\pi}\vert F(e^{i\theta})\vert^p d\theta\right]^{1/p} <\infty.
\end{equation}
In particular, $H^1$ is the Banach space of analytic functions
with integrable modulus on the boundary, and $H^2$ is the Hilbert space
of analytic functions with integrable modulus squared. We introduced already
  the class  $H^\infty$ of functions analytic and bounded in
 $\vert z\vert \le 1$, for which the $L^\infty$ norm was defined
in Eq.~(\ref{Hinf}). The classes $H^p$ and $H^q$ are said to be dual if the relation $1/p+1/q=1$ holds \cite{Duren}. It follows that $H^1$ and $H^\infty$ are dual to each other, while $H^2$ is dual to itself.

We now state the duality theorem which will be applied below. 
Let $h(\zeta)$ be a complex bounded function defined on the boundary of the unit disk (for simplicity we omit here the  dependence on $Q^2$).
Then the following equality holds (see Sec.~8.1 of Ref.~\cite{Duren}):
\begin{equation}\label{dual}
\min_{g\in H^\infty}\|g-h\|_{L^\infty}=\sup_{G\in S^1}
\frac{1}{2\pi}\left\vert\oint_{\vert \zeta\vert =1} G(\zeta)h(\zeta) d \zeta
\right\vert,
\end{equation}
where $S^1$ denotes the unit sphere of the Banach space $H^1$, \ie, the set of functions 
$G\in H^1$ which satisfy the condition $ \|G\|_{L^1}\leq 1$.

We note first that the equality~(\ref{dual}) is automatically satisfied  if $h$
 is the boundary value of an analytic 
function in the unit disk, since in this case
the minimal norm on the left-hand side is zero, and the right-hand side of Eq.~(\ref{dual}) vanishes too, by Cauchy's theorem.
 The nontrivial case corresponds  to a function $h$ 
 which  is not the
boundary value of a function analytic in
 $|z|< 1$ and which admits a general Fourier expansion containing both positive and negative-frequency terms. Restoring the $Q^2$ dependence, we expand $h(\zeta, Q^2)$ as
\beq\label{hpm}
h(\zeta, Q^2)=\sum\limits_{k=0}^\infty h_k(Q^2) \zeta^k + \sum\limits_{k=1}^\infty c_k(Q^2) \zeta^{-k}\,, 
\eeq
where  the $Q^2$-dependent Fourier coefficients $h_k$ and $c_k$ are real. The analytic continuation of the expansion (\ref{hpm}) from the boundary of the unit disk  $|z|=1$ to its interior $|z|<1$ will contain both an analytic part (the first sum) and a nonanalytic part (the second sum). Intuitively, we expect the minimum norm 
in Eq.~(\ref{dual}) to depend explicitly only on the nonanalytic part, \ie, on the coefficients $c_k$. 

In order to evaluate the supremum on the right-hand side of Eq.~(\ref{dual}) we
 use  a factorization theorem (see the proof of Theorem~3.15 in Ref.~\cite{Duren}) according to which
every function $G(z)$ belonging to the unit sphere $S^1$ of $H^1$ 
 can be written as
\begin{equation}\label{factor}
G(z)=w(z)f(z)\,,
\end{equation}
where the functions $w(z)$ and $f(z)$ belong to the unit sphere $S^2$ of $H^2$,
\ie, are analytic and satisfy the conditions
\begin{equation}\label{wGnorm}
\Vert w\Vert_{L^2} \leq 1\,,\quad \quad\quad\Vert f\Vert_{L^2}\leq 1\,.
\end{equation}
 Therefore, if one writes the Taylor expansions
 \begin{equation}\label{wf}
w(z)=\sum_{n=0}^\infty w_nz^n\,,\quad \quad f(z)= \sum_{m=0}^\infty f_m z^m\,,
\end{equation}
the coefficients satisfy  the conditions
\begin{equation}\label{wfl2}
\sum_{n=0}^\infty w_n^2\leq 1\,,\quad\quad \quad\sum_{m=0}^\infty f_m^2\leq 1\,.
\end{equation} 
After introducing the representation~(\ref{factor}) into Eq.~(\ref{dual}), we obtain the equivalent  relation
\begin{equation}\label{dual1}
\min_{g\in H^\infty}\|g-h\|_{L^\infty}=\sup_{w,f \in S^2}\left\vert \frac{1}{ 2\pi}
\oint\limits_{\vert \zeta\vert =1} w(\zeta)f(\zeta)h(\zeta) d\zeta\right\vert\,,
\end{equation}
where the supremum on the right-hand side is taken with respect to the functions 
$w$ and $f$ with
the properties mentioned in Eqs.~(\ref{wf}) and (\ref{wfl2}). By inserting
 into  Eq.~(\ref{dual1}) 
the expansions (\ref{wf})
we obtain, after a straightforward calculation
\begin{equation}\label{dual2}
\min_{g\in H^\infty}\|g-h\|_{L^\infty}=\sup_{\{w_n,f_m\}}
\left\vert\sum_{m,n=1}^\infty {\cal H}_{nm}w_{n-1}f_{m-1}\right\vert\,.
\end{equation}
The supremum is taken with respect to the sequences 
$w_n$ and $f_m$ subject to the condition (\ref{wfl2}), and the numbers
\beq\label{hank}
{\cal H}_{km}\equiv c_{k+m-1}(Q^2)\,, \quad k,m\geq 1,
\eeq
define a matrix ${\cal H}$ in terms of the  negative-frequency Fourier coefficients $c_k(Q^2)$  of the function $h$ expanded in Eq.~(\ref{hpm}), which are calculated as
\begin{equation}\label{four}
c_{k}(Q^2)=\frac{1}{2\pi }\int\limits_0^{2 \pi} e^{i k\theta}h(e^{i\theta}, Q^2 )  d\theta,\quad k\ge 1.
\end{equation} 
It is convenient to  write $c_k$ in an equivalent way as a contour integral along the boundary $|\zeta|=1$ of the unit disk:
\begin{equation}\label{four1}
c_{k}(Q^2) =\frac{1}{2\pi i}\oint\limits_{\vert \zeta\vert=1}\zeta^{k-1}h(\zeta, Q^2) d\zeta,\quad k\ge 1.
\ee

Matrices with elements defined as in (\ref{hank}) are called Hankel matrices \cite{Duren}. If  $w_{k-1}$ and $\sum_m{\cal H}_{km}f_{m-1}$ are viewed as the components of vectors
${\bf w}$ and ${\bf {\cal H}f}$,  the
absolute value of the sum in Eq.~(\ref{dual2}) can be written as
$\left\vert{\bf w}\cdot{\bf {\cal H}f}\right\vert$, and the Cauchy--Schwarz inequality implies that it satisfies
\beq
\label{CS}
\left\vert{\bf w}\cdot{\bf {\cal H}f}\right\vert\le\Vert{\bf w}\Vert_{L^2}
\Vert{\bf {\cal H}f}\Vert_{L^2}\le\Vert{\bf {\cal H}f}\Vert_{L^2}\,.
\eeq
Since Eq.~(\ref{CS}) is saturated for ${\bf w}\propto{\bf {\cal H}f}$, it follows that the
supremum in Eq.~(\ref{dual2}) is given by the $L^2$ norm of the matrix ${\bf {\cal H}}$.
The solution of the minimization problem~(\ref{eq:inf}) can then be written as 
\begin{equation}\label{delta0}
\mu_0(Q^2)=\Vert{\cal H}\Vert_{L^2}\equiv \Vert{\cal H}\Vert\,,
\end{equation}
where $\Vert{\cal H}\Vert$ is the spectral norm, given by 
 the square root of the greatest
 eigenvalue of the positive-semidefinite matrix ${\cal H}^\dagger {\cal H}$.

 In the numerical  calculations, as in previous works \cite{Caprini1, CGP2014, CaSa},  the matrix  ${\cal H}^\dagger {\cal H}$ has been truncated  at a finite order 
$m=n=N$, using the fact that for large $N$ the successive approximants tend toward the exact result (for a proof of the convergence see Appendix E of \cite{CiNe}).
 By the duality theorem, the initial
functional minimization problem~(\ref{eq:inf})
 was reduced to a rather simple numerical computation.

\vspace{0.5cm}

\end{document}